\documentclass [aps,prl,amssymb,amsmath,twocolumn,showpacs]{revtex4-2}

\usepackage{microtype}
\usepackage{amsmath}
\usepackage{amssymb}
\usepackage{amsthm}
\usepackage{mathrsfs}
\usepackage{graphicx}
\usepackage{verbatim}
\usepackage{bm}
\usepackage[dvipsnames]{xcolor}

\usepackage{graphicx,epsfig,amsfonts,amssymb}

\usepackage[colorlinks,
linkcolor=blue,
anchorcolor=blue,
urlcolor=blue,
citecolor=blue
]{hyperref}

\usepackage{bm}
\usepackage{times}
\usepackage{lipsum}

\usepackage[all]{xy}

\newcommand{\nn}{\nonumber}

\newcommand{\br}{\mathbf r}

\newcommand{\be}{\begin{eqnarray}}
\newcommand{\ee}{\end{eqnarray}}
\newcommand{\la}{\langle}
\newcommand{\ra}{\rangle}
\newcommand{\rar}{\rightarrow}

\begin{document}
	\preprint{APS/123-QED}
	
	\title{Statistical Analysis of Magnetic Domain Wall Dynamics to Quantify  Dzyaloshinskii-Moriya Interaction}
	\author{Xinwei Shi}
	\author{Chen Sun}
	\author{Fuxiang Li}
	\email{fuxiangli@hnu.edu.cn}
	\affiliation{School of Physics and Electronics, Hunan University, Changsha 410082, China}
	
	\begin{abstract}
		We utilize  statistical tools to analyze the magnetic domain wall dynamics in a nanostrip, which can quantify the magnitude and reveal the effects of interfacial Dzyaloshinskii-Mariya interaction. We find that there exist two peaks in the velocity frequency spectrum, the magnitude ratio of which can be used to determine the DMI strength. Our approach is validated using a collective-coordinate model, and is demonstrated to be robust  against thermal noise and material impurities. Moreover, third-order cumulant and third-order time-dependent correlation function of velocity are  calculated and yield valuable information regarding the asymmetry induced by DMI. Our findings offer  novel and efficient analysis tools to understand  physical process of domain wall dynamics under DMI and exotic magnetic phenomena.
	\end{abstract}
	\maketitle
	
	%

	{\it Introduction.~}Dzyaloshinskii-Moriya Interaction (DMI) plays a fundamental role in the stabilization of exotic spin structures, such as spin spirals \cite{wang2017controlling,von2014interface}, chiral domain walls \cite{li2019chiral,brataas2013chiral}, and magnetic skyrmions \cite{fert2017magnetic,finocchio2016magnetic,zhang2020skyrmion}. 
	These spin structures can be driven by electric current or external magnetic field with high level of efficiency and thus are particularly attractive as promising information carriers for future spintronics technologies \cite{wolf2001spintronics,fert2017magnetic}. Therefore, accurately and reliably quantifying the DMI in magnetic materials is of utmost importance in identifying potential materials for spintronic-based devices.  
	
	One particularly interesting and commonly studied magnetic system is the thin magnetic film with perpendicular magnetic anisotropy due to its fast domain-wall motion driven by  electric current or magnetic field \cite{li2010current,shepley2015modification,koyama2008control,pham2016very}. In these multilayers consisting of ultrathin ferromagnetic film in contact with heavey metal, interfacial DMI can be induced on the interface due to the broken inversion symmetry and the large spin-orbit coupling of the heavy metal atoms \cite{carpentieri2015topological,lin2020perpendicular}. 
	It has been demonstrated that DMI strongly affects  the domain wall internal spin texture,  domain wall dynamics, and spin wave propagation  \cite{mougin2007domain,thiaville2012dynamics}. In return,
	by exploiting the unique properties and peculiar phenomena induced by DMI, such as Walker breakdown in field-driven domain wall motion and non-reciprocity in spin-wave propagation, different experimental techniques have been developed to quantify DMI strength in combination with simple analytical theoretical model \cite{kuepferling2023measuring}. For example,  using the Brillouin light spectroscopy, the DMI strength can be obtained by monitoring the nonreciprocal propagation in the Damon-Eshbach geometry \cite{zakeri2010asymmetric,moon2013spin}.  In the creep regime of domain wall motion, the asymmetric expansion of magnetic bubble under in-plane magnetic field and in the presence of DMI, can also be utilized as a measure of DMI strength \cite{kuepferling2020measuring,garcia2021magnetic}. 
Most of the experimental techniques, however,  require either high-precision imaging of domain wall structure or high quality of magnetic material \cite{kuepferling2023measuring}. 	
	

	
	Here, we  introduce statistical tools to analyze the domain wall dynamics in a magnetic nanostrip in the presence of DMI. We show that, rather than simple averaging of  instantaneous velocity \cite{soucaille2016probing,metaxas2007creep}, statistical analysis reveals more detailed information concealed in the noisy data of magnetic domain wall dynamics. By fast Fourier transformation of time-varying domain wall velocity, one observes the emergence of two peaks in the velocity frequency spectrum in the presence of interfacial DMI. By combining the micromagnetic simulation and analytical analysis based on the collective-coordinate model, we show that, in the precessional regime,  the magnitude ratio of the low frequency mode to the high frequency mode in the velocity frequency spectrum is linearly proportional to the DMI strength, but is independent of external magnetic field.   This ratio can thus be utilized to experimentally quantify the DMI strength in ferromagnetic films.
	Further, this method of velocity frequency spectrum  is demonstrated to be robust  even in the presence of external noise and magnetic pinning disorder, which is crucial for accurately measuring the strength of DMI in real materials. Moreover, third-order cumulant and third-order time-dependent correlation function of velocity are  calculated and shown to yield valuable information regarding the asymmetry induced by DMI. 
	The proposed statistical analysis provides a novel and robust method in quantifying micromagnetic parameters and uncovers  more detailed information of DMI.


	
	{\it Model and setup.~} We start from considering the field driven domain wall dynamics of a thin magnetic film with perpendicular magnetic anisotropy. The film is patterned into a long strip which provides an ideal setup to study the domain wall propagation. We use the GPU accelerated micromagnetic simulation program MuMax3 to simulate the DW dynamics \cite{vansteenkiste2014design}. This program solves the space and time dependent reduced magnetization $\bm{m}(\br, t)$   in the Landau-Lifshitz-Gilbert (LLG) equation,
	\begin{equation}
		\dfrac{\partial \bm{ {m}}}{\partial t}=-\gamma \bm{ {m}} \times {\bm H}_{\rm eff}+\alpha \bm{ {m}} \times\dfrac{\partial \bm{ {m}}}{\partial t}.
	\end{equation}
	Here, $\gamma$ is the gyromagnetic ratio, $\alpha$ the dimensionless damping parameter, and $\boldsymbol{H}_{\rm eff}$ the effective field consisting of  externally applied field, magnetostatic field, Heisenberg exchange field, as well as the anisotropy field. In the simulation, we choose thickness $L_{z}$ = 3 nm, width $L_{y}$ = 20 nm and length $L_{x}$ = 4096 nm. Micromagnetic parameters are chosen to be the typical experimental values \cite{metaxas2007creep}: saturation magnetization $M_{s}= 9.1\times10^{5}$ A/m, exchange stiffness $A_{\rm ex} =1.4\times10^{-11}$ J/m, first order uniaxial anisotropy constant $K_{u} = 8.4\times10^{5}$ J/m$^{3}$, and damping parameter $\alpha= 0.27$. The DWs width  is roughly given by $\Delta=\sqrt{A_{\rm ex}/K_{\rm eff}}$, where $K_{\rm eff}$ is the effective anisotropy, which includes the magnetocrystalline
	anisotropy $K_{u}$ and the shape anisotropy.
	The discretization cell dimensions are $d_{x} = d_{y} =2$ nm, $d_{z}  =0.5$ nm, smaller than the exchange length $L_{\rm ex}=\sqrt{A_{\rm ex}/(\mu_{0} M_s^2)}\approx 3.7$ nm \cite{abo2013definition}. Periodic boundary conditions are used in the $y$ direction to avoid boundary effects \cite{herranen2019barkhausen}. The LLG equation is then solved using the Dormand-Prince solver (RK45) with an adaptive time step.
	
\begin{figure}
		\centering
		\includegraphics[width=0.5\textwidth]{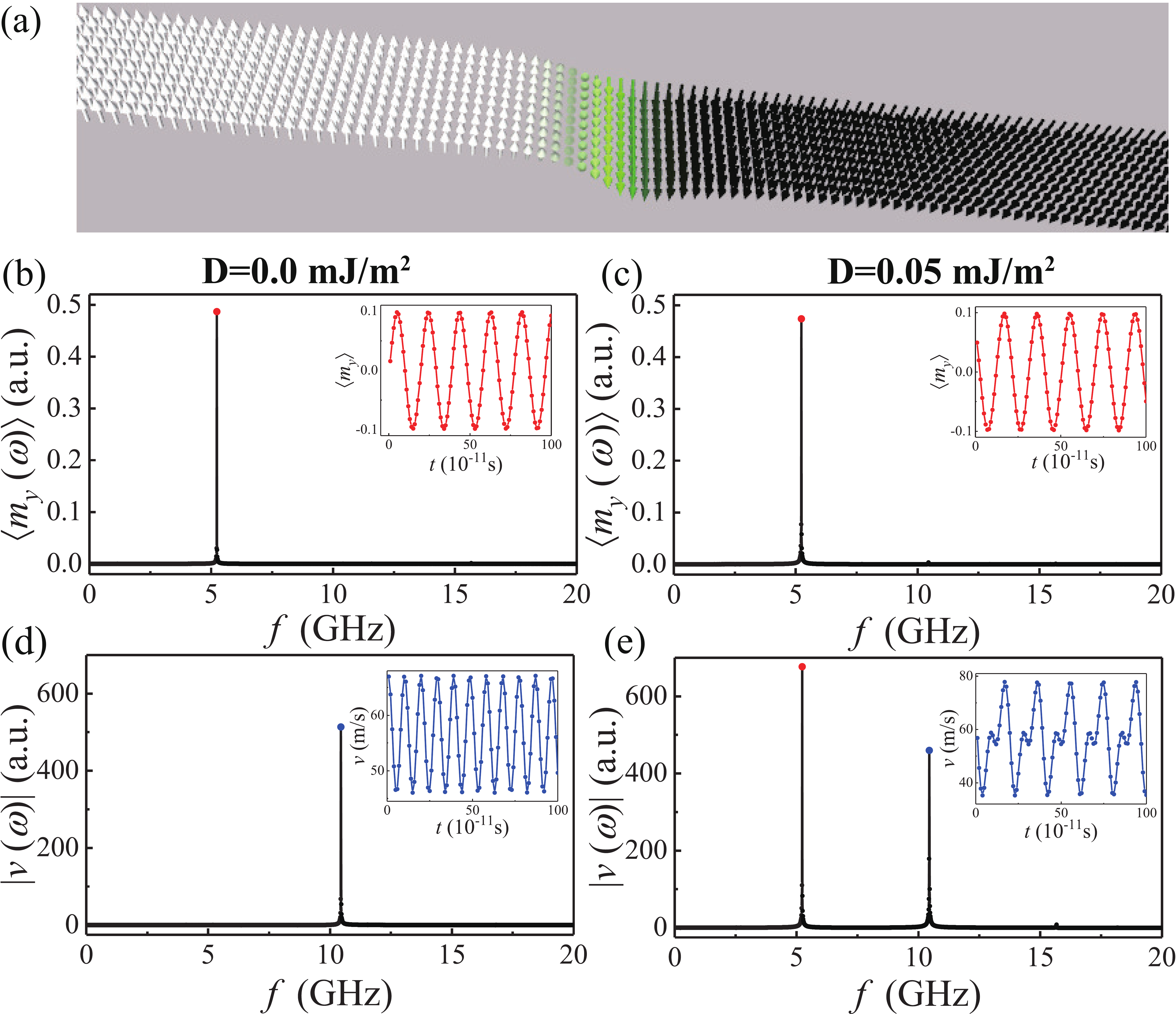}\\
		\caption{(a) Schematic illustration of field-driven DW motion in  magnetic nanostrip. (b) and (c) Frequency spectra of  $\langle m_{y}(t) \rangle$ without and with DMI, respectively.  (d) and (e) Frequency spectra of  $v(t)$ without and with DMI, respectively.  The red and blue points correspond to the LFM and HFM peaks respectively.	Inset: $\langle m_{y} \rangle (t)$ (red dotted line) and $v_{\rm DW}(t)$ (blue dotted line) signals. External field $B_{\rm{ext}}$=200 mT.}\label{fig1}
	\end{figure}

	The system is initialized in a configuration with two antiparallel out-of-plane (up  and down) domains separated by a DW with an internal magnetization in the negative $y$-direction, as shown in Fig.~\ref{fig1}(a). The behavior of DW in response to an external perpendicular magnetic field $B_{\rm{ext}}$ is featured with unique, nonlinear dynamics that has been understood very well \cite{thiaville2012dynamics}. Before the Walker breakdown, the averaged velocity increases linearly with external magnetic field, and after that, the velocity  suddenly drops, signaling the onset of precession of the DW \cite{schryer1974motion}. In the Supplemental Material \cite{sm} the DW dynamics driven by electric current is studied. In the simulation, one can record the information of time varying quantities $m_{x, y, z}(t)$, from which the DW dynamics can be further investigated. The average $\langle m_{i} \rangle$ ($i = x, y, z$) is taken over a range extending 20 discretization cells around the DW.  The instantaneous domain wall velocity is obtained by $v(t) \propto d \langle m_{z} \rangle /{d t} $ \cite{vandermeulen2015collective}. 
	
	{\it Velocity frequency spectum.} In the precessional regime with zero DMI strength $D=0$, one observes that, both the $\langle m_{x,y}(t) \rangle$ and $v(t)$  oscillate periodically with time, as shown in the inset of Fig.~\ref{fig1}(b) and (d). However, they have different oscillating frequency. 
	The difference can be seen more clearly from the frequency spectrum of  $\langle m_{y}(t) \rangle$ and $v(t)$.  After a long time interval $T$,  one makes  the fast Fourier transform of the instantaneous velocity $v(t)$:
	\begin{equation}
		v(\omega)=\dfrac{1}{\sqrt{T_{m}}} \int_{0}^{T_{m}} e ^{i\omega t}  v(t ) dt. 
	\end{equation}
	Similar Fourier transform is also performed for $\langle m_{y}(t) \rangle $. 
	In Fig.~\ref{fig1}(b) and (d), one plots the magnitude of Fourier transform $|\la m_y(\omega)\ra$ and  $|v(\omega)|$ for the case of $D=0$. It is shown that, there is only a single peak in the frequency spectrum of $m_y(t)$ and $v(t)$. However, the peak frequency of velocity is doubled compared with  that of $m_y$. 
	
	Interesting phenomenon appears when one introduces nonzero DMI.  In the presence of  DMI, the time-varying signal $\langle m_{y}(t) \rangle$ is almost the same as  $D$ = 0, and the oscillation frequency remains unchanged (inset of Fig.~\ref{fig1}(c)). However, the behavior of velocity signal $v(t)$ is radically changed, as shown in the inset of Fig.~\ref{fig1}(e).   
	Figures ~\ref{fig1}(c) and (e) plot  the  frequency spectra of $\langle m_{y} \rangle$ and $v(t)$ with nonzero DMI. While there is only one peak in the frequency spectrum of $\langle m_{y} \rangle$, there  emerge two spectral peaks with  different magnitudes in $|v(\omega)|$.  
	For convenience, we label the two oscillatory modes as the low-frequency mode (LFM) and high-frequency mode (HFM), respectively.

	\begin{figure}
		\centering
		\includegraphics[width=0.5\textwidth]{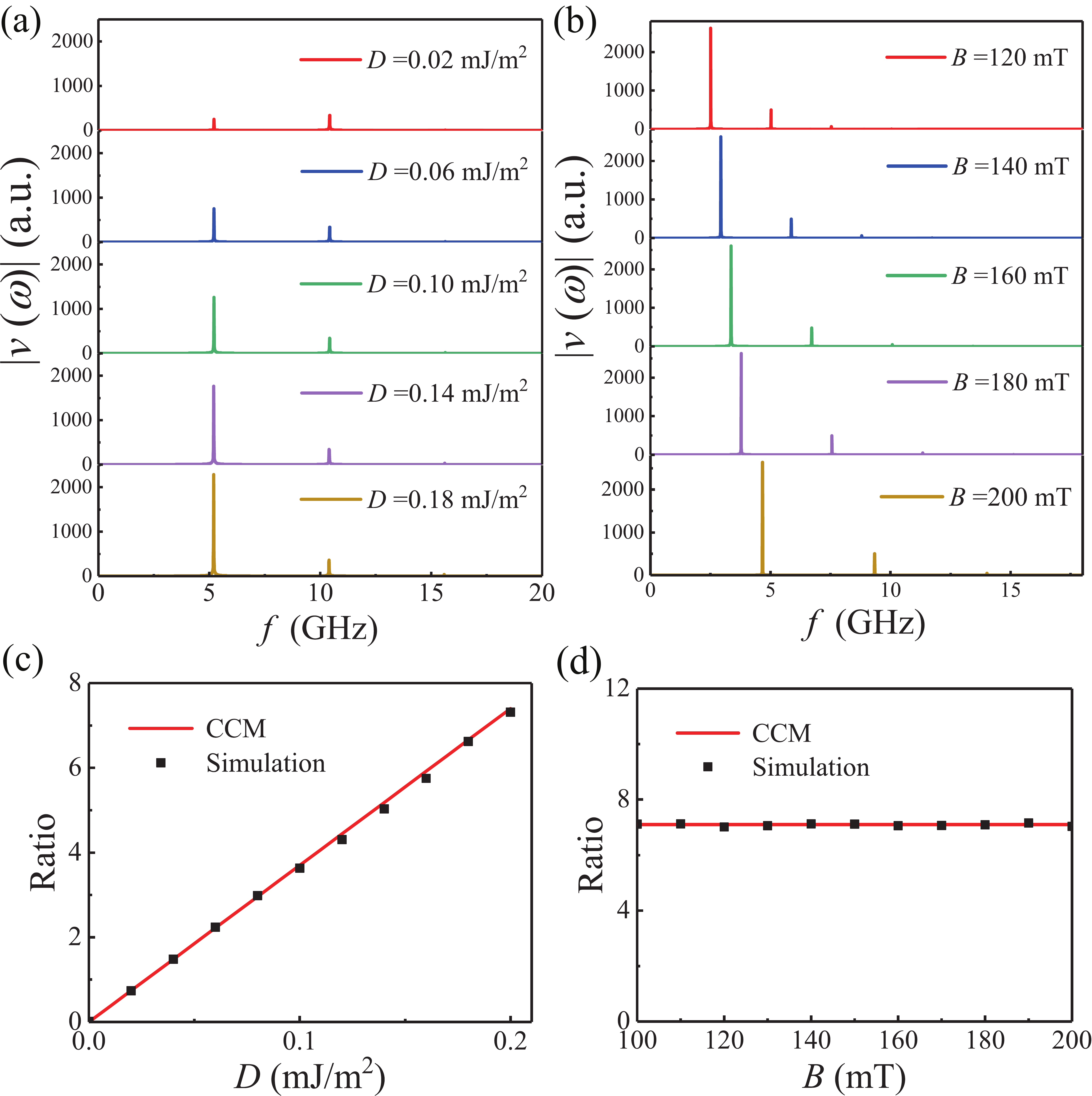}\\
		\caption{Frequency spectrum of DW velocity $v(t)$ for different values of  $D$ (a) and $B_{\rm ext}$ (b). Dependence of the ratio of  LFM to HFM  on  $D$ (c) and $B_{\rm ext}$ (d). 
			Squares are results from micromagnetic simulation and the solid line  from the  CCM.}\label{fig2}
	\end{figure}
	
	We further investigate the influence of DMI strength $D$ on the two emerging oscillating peaks in the velocity frequency spectrum.  We first  fix  the external magnetic field $B_{\rm ext}$ and gradually increase the DMI strength $D$. As shown in Fig.~\ref{fig2}(a), there are always two peaks in the velocity frequency spectrum with the frequencies keeping unchanged. However, with increasing DMI strength, the magnitude of the LFM peak increases, while the magnitude of HFM peak remains  unchanged. If we extract the LFM/HFM ratio from the frequency spectral data and plot them as a function of DMI intensity  (Fig.~\ref{fig2}(c)), one can see that the ratio  increases  linearly with the DMI strength $D$. 
	We also  study the velocity frequency spectrum  under different external magnetic fields $B_{\rm ext}$ when the DMI strength is fixed, as shown in Fig.~\ref{fig2}(b) and (d). Even though the oscillating frequency increases with increasing $B_{ext}$, the magnitude ratio of LFM/HMF remains unchanged as long as the DMI strength $D$ is fixed.  
	
	This result confirms that the magnitude ratio of the two peaks in velocity frequency spectrum is solely dependent on DMI, and thus can be utilized as a measure of DMI strength. Later we will show that this frequency spectrum analysis remains robust in the presence of  external noise and pinning.

	{\it Collective-coordinate model.}
	To understand the underling physics of the above results obtained from the micromagnetic simulation, we resort to the collective-coordinate model that can provide semi-analytical solution of domain-wall dynamics.  
	Assuming that the DW maintains as a rigid object with the width remaining constant, the DW dynamics can be described by two independent variables, the DW position $q$ and its conjugate momentum, the DW magnetization angle $\varphi$. 
	The 1D collective-coordinate model (CCM) reads \cite{tretiakov2010current,sun2022field,jung2008current,boulle2013domain,martinez2014current,conte2015role,lucassen2009current}:
	\be
	&&v(t) \equiv \frac{dq}{dt}= \gamma' \Delta (\alpha H_a - H_K \sin2\varphi + H_D \sin \varphi),   \label{eqvt}\\
	&&  \frac{d\varphi}{dt} =\gamma' (H_a + \alpha H_K \sin2\varphi -\alpha H_D \sin\varphi), \label{eqphi1}
	\ee
	with $\gamma'= \dfrac{\gamma }{\alpha^2+1}$.
	Here, $H_{K}=2K /(\mu_{0} M_{s})$ with $K$ being the effective anisotropy energy, $H_a$ is the external magnetic field applied along the easy axis, and $H_{D}=\pi D /(2\mu_{0} M_{s} \Delta)$.  
	Despite its simplicity, the CCM provides a quite accurate description of DW motion in a nanowire. It is also easy to generalize the equations to the cases with external thermal noise or pinning disorder, and to the system driven by electric current \cite{sm}. Now we use Eqs.\eqref{eqvt} and \eqref{eqphi1} to explain the emergence of two peaks in the velocity frequency spectrum in the presence of nonzero DMI. 
	
	Note that without external noise or disorder, the average magnetization $\la m_y (t)\ra$ is almost a sinusoidal, as depicted in the inset of Fig.~\ref{fig1}(b) and (c). Since $\la m_y(t) \ra$ is proportional to $\sin\varphi$, this means that $\sin\varphi$ is also an almost perfect sinusoidal with only a slight deviation. 
	Actually, one can see this point simply from   Eq.~\eqref{eqphi1}, which can be analytically solved if one of the two terms containing $H_K$ and $H_D$ vanishes. If $H_K=0$, the solution of $\varphi(t)$ is given by: 
	\be
	\tan\frac{\varphi}{2} = \frac{\sin(t'/2)}{\cos(t'/2+\theta)}
	\ee
	with  $t'=\gamma' H_a \sqrt{1-(\alpha H_D/H_a)^2}$ and $\theta= -\arcsin(\alpha H_D/H_a)$. The solution shows that if $\theta\ll1$, i,e., $\alpha H_D\ll H_a$, then $\sin\varphi$ is very close to a sine function of time $t$, and the DW magnetization angle $\varphi(t)$ can be considered to increase almost linearly with time $t$.  Indeed, direct calculations of the Fourier coefficients of $\sin\varphi$ show that, small deviation from linearity only results in the appearance of peaks at $n$-th harmonic frequency, but with small amplitude that is of the order  $\theta^{n-1}$ (see SM for detailed discussion \cite{sm}). Similar discussion can be made for the case with nonzero $H_K$. 
	
	Under this consideration, one can qualitatively understand the previously observed phenomena from micromagnetic simulations. First,  for the case of $H_D=0$, the time dependent part of velocity in Eq.~(\ref{eqvt}) is proportional to $\sin 2\varphi$, leading to a peak at the second harmonic frequency in the frequency spectrum. As one increases the DMI strength $D$, and thus $H_D$, an additional component proportional to $\sin \varphi$ appears, and gives rise to a LFM in the velocity frequency spectrum. More importantly,  in the limit that $\varphi$ is almost linear in time $t$, the magnitude ratio of  LFM to HFM is given from Eq.~\eqref{eqvt} by: 
	\begin{equation}
		\eta = \frac{H_D}{ H_K} = \frac{\pi D}{4\Delta K} \label{eta}
	\end{equation}
	Therefore, we conclude that the magnitude of DMI strength $D$ can be easily obtained once we get the ratio $\eta$. The accuracy of this approach is guaranteed by $\alpha H_D/ H_a \ll 1$ and $\alpha H_K/ H_a \ll 1$, which can be easily achieved by increasing external field $H_a$.

	\begin{figure}
		\centering
		\includegraphics[width=0.5\textwidth]{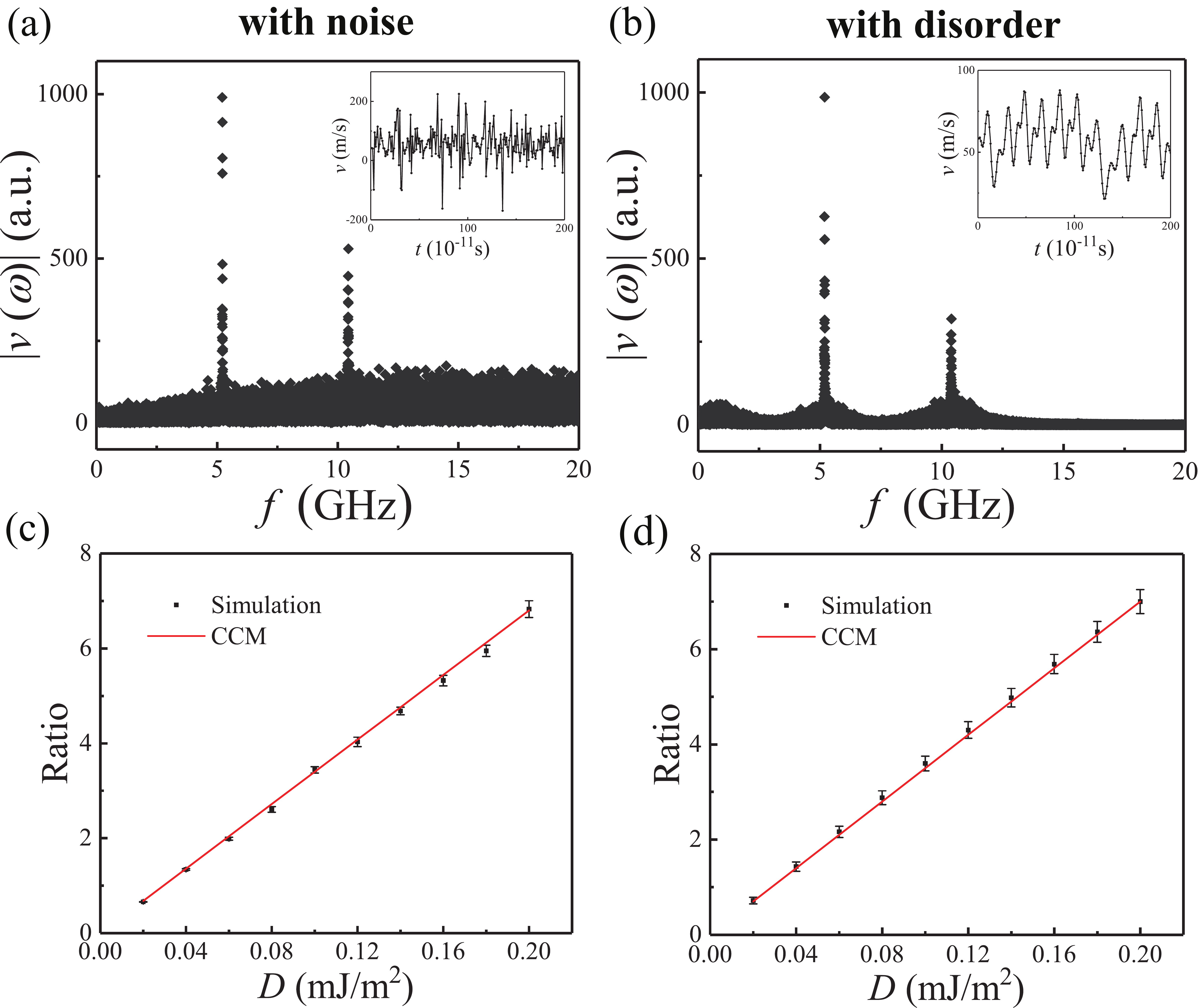}\\
		\caption{Velocity frequency spectrum and magnitude ratio of LFM/HFM in the presence of external noise and pinning disorder. (a) and (b) are the velocity frequency spectra with external noise and pinning disorder, respectively. 
			Parameters: $D$=0.05 mJ/m$^{2}$, $R=10$ mT, $r=0.03$.  The insets are the corresponding velocity signals. (c) and (d) are the corresponding magnitude ratio vs. DMI strength $D$.   The error bars on the simulated data correspond to the uncertainties in the averaged magnitude of the frequency spectrum.}
		\label{fig3}
	\end{figure}
	
	{\it External noise and pinning disorder.}
	We now proceed to verify the robustness of the frequency spectrum method and the accuracy of Eq.~\eqref{eta} in quantifying DMI strength in the presence of external noise and pinning disorder. In this case, the time-varying velocity $v(t)$ is no longer a well-defined sinusoidal. However, one can repeat the measurement many times during a short time interval, and take the average of fast Fourier transform $v(\omega)$.  This method of data processing eliminates the effects of noise and disorder and provide the true information of DMI strength.

	We first investigate the effect of external noise on the velocity frequency spectrum.  Without loss of generality, we consider the Gaussian white noise \cite{jovkovic2021effects,spasojevic2022interplay,gopalakrishnan2015effect,janicevic2021scaling}. This type of noise commonly originates from experimental apparatus \cite{bittel1969noise}, such as detectors \cite{kaisar2021modeling}, amplifiers \cite{netzer1981design}, and ambient electromagnetic interference \cite{jovkovic2021effects}. 
	In the micromagnetic simulation, the effect of external noise can be introduced by the   random field vector $\bm{h}_j$ on each site $j$, giving rise to an additional Zeeman energy $-M_s \sum_j {\bm h}_j \cdot{\bm m}_j$.  Taking the average  over all possible random field configurations, one has $\la {\bm h}_j \ra =0$  and  $\langle h_{i \alpha} h_{j \beta}\rangle = R^{2}\delta _{i,j}\delta _{\alpha, \beta}$ where $\delta _{i,j}$ is the Kronecker delta function, $\alpha, \beta = x, y, z$, and $R$ measures the disorder strength, i.e. the standard deviation of the employed random field distribution. In this paper, we adopt the   Gaussian distribution \cite{jovkovic2021effects} $\rho(h)=\exp(-h^{2}/2 R^{2})/(\sqrt{2\pi}R)$. 
	Extensive simulations are conducted on system sizes up to $L_x$ = 2048 for sufficiently strong noise $R= 10$ mT.  The total time of simulation is $10^{-5}$s, with $10^6$ data points taken. 
	To obtain the frequency spectrum, one divides the total time series into relatively short time intervals with each time interval consisting of $10^4$ data points. One then makes a Fourier transform of each short time interval and take the average of the Fourier transform $|v(\omega)|$ over all the time intervals.  
	Fig.~\ref{fig3}(a) plots the  velocity frequency spectrum and the inset is a snapshot of time-varying velocity $v(t)$ with $200$ data points. One can see that the velocity is strongly disturbed, and is no longer a well defined sinusoidal. It is therefore hard to tell the information of DW dynamics simply from the profile of velocity. Nevertheless, after a sufficiently long time average, one can still obtain two peaks identified as HFM and LFM in the velocity frequency spectrum, similar to the clean system. Here, the only difference is that, due to the external white noise, there appears a non-zero  background. Fig.~\ref{fig3}(c) plots the magnitude ratio of LFM to HFM as a function of DMI strength $D$ when external noise is added.
	
	We now consider the effect of intrinsic material defects in a real ferromagnetic nanostrip which can significantly impact the behavior of domain walls \cite{tanigawa2008dynamical,caballero2018magnetic,jeudy2018pinning}.  These defects create potential wells in the micromagnetic energy landscape, which can be characterized by the saturation magnetization and anisotropy between grains \cite{gaunt1983ferromagnetic,klaui2005direct,moretti2017dynamical}. 
	For thin films with thicknesses of only a few atoms, a natural source of disorder is given by thickness fluctuations of the film \cite{zapperi1998dynamics}. In order to account for the effect of quenched disorder, we construct “grains” of linear size 20 nm (defining the disorder correlation length) by Voronoi tessellation. Each grain has a normally distributed random thickness $t_{G}=h+{\rm Norm}(0, r)  h$, with $r$ the relative magnitude of the grain-to-grain thickness variations and $h$ the mean thickness of the sample. ${\rm Norm}(0, r)$ denotes a normal distribution function with mean $0$ and standard deviation $r$.   These thickness fluctuations are then modeled using an approach proposed in Ref. \cite{herranen2019barkhausen}, by modulating the saturation magnetization and anisotropy constant according to $M_{s}^{G}=M_{s}t_{G}/h$ and $K_{u}^{G}=K_{u} t_{G}/h$. 
	Fig.~\ref{fig3} (b) and (d) present the velocity $v(t)$ over time, its frequency spectrum and the ratio LFM/HFM as a function of DMI strength $D$ in the presence of disorder. One can see that the profile of velocity $v(t)$ is also perturbed leading to a nonzero background  and larger peak width in the frequency spectrum. However, the ratio $\eta$  still linearly increases with DMI strength $D$, verifying the robustness of the frequency spectrum method. 
	
	
	\begin{figure}
		\centering
		\includegraphics[width=0.5\textwidth]{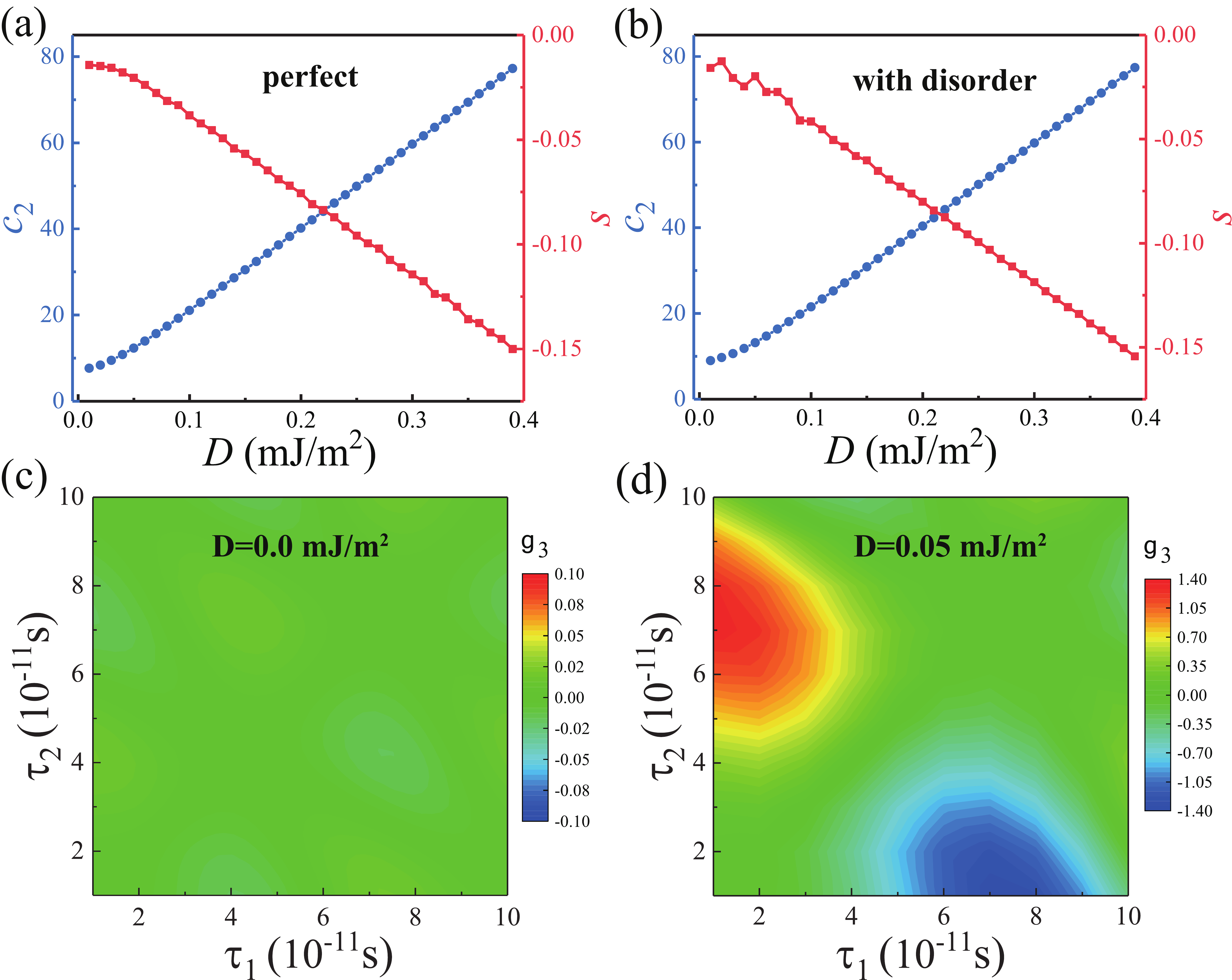}\\
		\caption{Skewness and third-order correlator. (a) and (b) are the standard variance (blue) and skewness (red) in the clean system and in system with disorder, respectively.   (c) and (d) are the normalized third-order correlator $g_3(\tau_1, \tau_2)$ in the clean system for $D=0$ mJ/m$^2$ and $D=0.05$ mH/m$^2$, respectively.  }
		\label{fig4}
	\end{figure}
	
	{\it Skewness and third-order correlation.}
	While  the average, variance, and second-order correlation of physical quantities can provide much information of domain wall dynamics,  more detailed information is actually concealed in the seemingly useless noisy data. Applying statistical methods on the noisy data, such as the distribution of the fluctuations, their moments, and autocorrelation function, can yield a wealth of information about the underlying dynamics \cite{qian2004fluorescence,chen2023chaotic}. Here, we show that considering the third-order cumulant and third-order time-correlation function can reveal the intrinsic asymmetry induced by DMI.
	In full counting statistics, one can consider all orders of cumulants of a physical quantity, here $\delta v(t) = v(t) - \la v \ra$: 
	\be
	c_n= \la \delta v^n \ra,
	\ee 
	with the average taking over a long period of time.
	By definition, the first order $c_1=0$. The second order cumulant, i.e., the variance, describes the magnitude of fluctuation. The third-order cumulant $c_3$  is a measure of the asymmetry of the probability distribution of a real-valued random variable about its mean. To quantify the asymmetry, one can further define the  skewness as:
	\be
	s={c_3}/{c_2^{3/2}}
	\ee
	to renormalize the third-order cumulant by the variance. This skewness   value can be positive, zero, negative, or undefined. In our case, as shown in Fig.~\ref{fig4}(a) and (b), one can see that  the skewness is always negative, and  the magnitude of skewness increases in a roughly linear behavior with $D$.  
	
	More detailed dynamical information can be revealed by studying the time-dependent correlation function \cite{mccoy1983time,schutz1993time}. Here we consider the second-order and third-order correlation function of velocity $v(t)$: 
	\be
	&&g_2(\tau) =\la \delta v(0) \delta v(\tau) \ra, \\
	&&g_3(\tau_1, \tau_2) =\la \delta v(0) \delta v(\tau_1) \delta v(\tau_1 + \tau_2)  \ra /c_2^{3/2}.
	\ee
	The Fourier transformation of second-order correlator $g_2(\tau)$ provides similar information as the frequency spectrum of $v(\omega)$. The third-order correlator $g_3$ normalized by the variance $c_2$, depends on two time variables and is expected to provide valuable information. In Fig.~\ref{fig4}(c) and (d), $g_3(\tau_1, \tau_2)$ is plotted for the cases with $D=0$ and $D\neq 0$, respectively. While $g_3$ is very small for the case with $D=0$ (note the scale of color bar), it becomes nonzero at finite time for the case with $D\neq 0$. More importantly, it displays a strong asymmetry with respect to $\tau_1$ and $\tau_2$, which should be a strong indication of the effect of DMI.

	{\it Discussion and conclusion.~} 
	We demonstrate that statistical analysis of the domain wall dynamics can be a powerful tool in quantifying DMI strength and can provide more detailed information of the effects induced by DMI. Compared with other experimental techniques in measuring DMI strength, the velocity frequency spectrum method is robust against external noise and pinning disorder, and thus doesn't require a clean system in real experiment. This method doesn't need high-precision imaging of domain-wall internal structure. It only requires  the recording of time-varying total out-of-plane magnetization $\la m_z(t) \ra$, which can be accomplished by the time resolved magneto-optical imaging technique based on magneto-optical Kerr or Faraday effect. 
	Moreover, third-order cumulant and third-order time-dependent correlation function of velocity are  calculated and shown to yield valuable information regarding the asymmetry induced by DMI. 
	Our findings offer a comprehensive understanding of the dynamics of domain walls in the presence of DMI and provide important insights for the development of novel DW-based devices.
	
	{\it Acknowledgments.}	This work was supported by the National Key Research and Development Program of the Ministry of Science and Technology (Grant No. 2021YFA1200700), the National Natural Science Foundation of China (Grant No. 11905054, No. 12275075, No. 12105094) and the Fundamental Research Funds for the Central Universities from China.

	

\newpage
\section{Supplemental Materials}

\subsection{Domain wall dynamics before the walker breakdown}
\begin{figure}[h]
	\centering
	\includegraphics[width=0.4\textwidth]{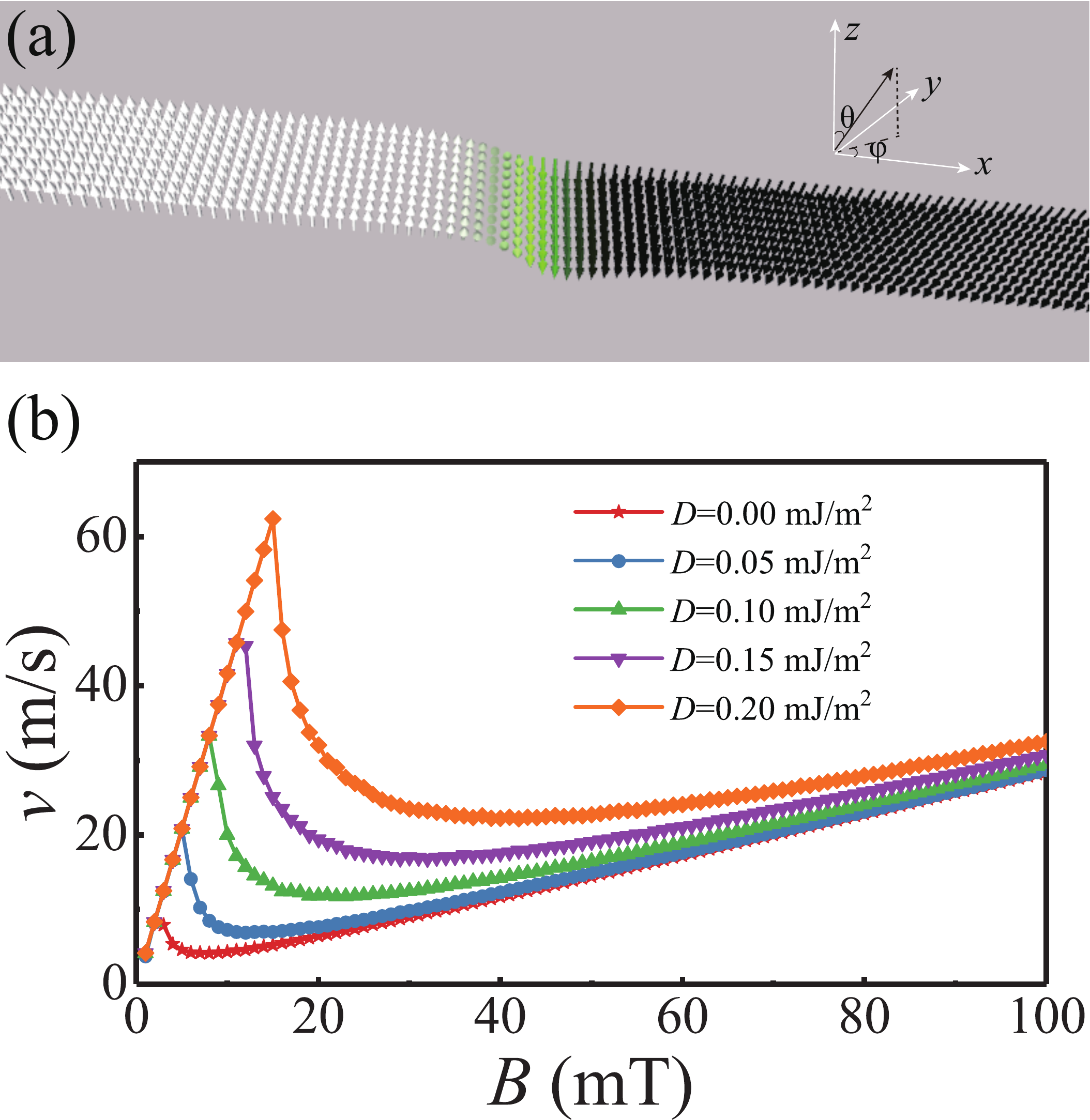}\\
	\caption{(a) Illustration of the domain wall structure in the magnetic nanostrip. (b) Averaged velocity as a function of external magnetic field for different strengths of DMI. }
	\label{figDW}
\end{figure}

Here we provide the results of domain wall velocity as a function of external magnetic field $B$ for different values of DMI strength in Fig.~\ref{figDW}. One can see that the DW velocity first linearly increases with increasing magnetic field until the Walker breakdown field after which the velocity decreases and then increases again. The presence of DMI can significantly increase the Walker breakdown field.

\begin{figure}[h]
	\centering
	\includegraphics[width=0.4\textwidth]{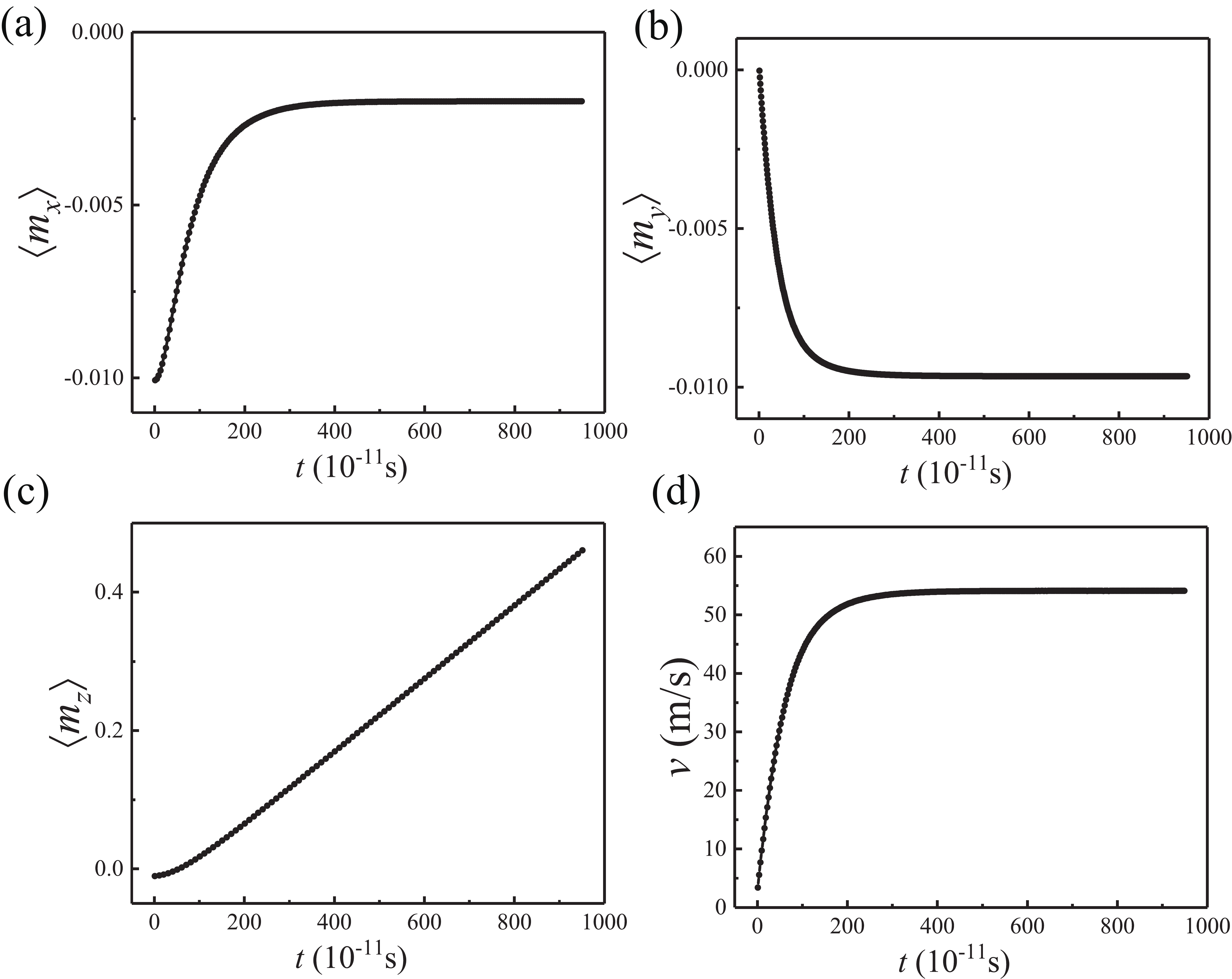}\\
	\caption{Domain wall dynamics before the Walker breakdown. (a-d) are the time-dependent averaged magnetization $\la m_{x, y, z} (t)\ra$ and the instantaneous velocity $v(t)$, respectively. }
	\label{figstationary}
\end{figure}

 In the main article, our study has been focused on the precessional regime. In this section, we also provide the dynamics in the stationary regime before the walker breakdown. In Fig.~\ref{figstationary} we plot the time dependent averaged magnetization $\la m_{x, y, z} (t)\ra$ and the velocity $v(t)$. As expected, there is no oscillation in this stationary regime and our approach analyzing the DMI is not applicable. 

\subsection{Exact solution of $\varphi$ in the collective coordinate model and its Fourier coefficients}
From the equations of motion in the collective coordinate model (CCM):
\be
		&&\dfrac{d\varphi}{dt}+\alpha\dfrac{1}{\Delta }\dfrac{dq}{dt}=\gamma H_{a}, \label{eqphi} \\
		&&\dfrac{1}{\gamma} \Bigg(\alpha \dfrac{d\varphi}{dt} - \dfrac{1}{\Delta }\dfrac{dq}{dt} \Bigg)=H_{K} \sin 2 \varphi -H_{D} \sin \varphi, \label{eqq}
		\ee
one can obtain the equation of $\varphi$ :
\be
\frac{d\varphi}{dt} =\frac{\gamma}{1+\alpha^2} (H_a + \alpha H_K \sin2\varphi -\alpha H_D \sin\phi)
\ee
We will show that if $\alpha H_K \ll H_a$ and $\alpha H_D \ll H_a$, our approach has high precision in quantifying the DMI strength. 
This equation is analytically solvable if $H_K=0$ or $H_D=0$. We first consider the case with $H_K=0$, and study the following reduced equation: 
\be
\varphi_t = 1+ a \sin\varphi \label{eqphi1}
\ee 
which can be obtained by making a time rescale $t\rar (1+\alpha^2)t/(\gamma H_a)$, and $a=-\alpha H_D/H_a$. 
Eq.~(\ref{eqphi1}) can be solved analytically. Its solution is given by
\be
\tan\frac{\varphi}{2} =\frac{\sin(t'/2)}{\cos(t'/2 + \theta)}
\ee
with $\theta = \arcsin a$ and $t'=\sqrt{1-a^2} t$.
Therefore,
\be
\sin\varphi = \frac{\sin(t'+\theta) - \sin\theta}{1-\sin(t'+\theta) \sin\theta}
\ee
and 
\be
\sin(2\varphi)=\frac{4\cos\theta \cos(t'+\theta)(\sin(t'+\theta)-\sin\theta)}{(1-\sin(t'+\theta) \sin\theta)^2}
\ee

One can see that $\sin\varphi$ is a periodic function of rescaled time $t'$ with period $2\pi$.
Now one can calculate the Fourier series of $\sin\varphi$
 and $\sin2\varphi$. For $\sin\varphi$, the Fourier coefficients are given by
 \be
 A^{(1)}_n = \int \frac{dt'}{2\pi} \sin(\varphi(t')) e^{i n t'} 
  \ee
  with $n=1, 2, 3, \ldots$.
This  integral can be performed in the complex plane by introducing $z=e^{i t'}$, and the integration contour is  along the unit circle. Using residual theorem, this integration can be performed exactly.
\be
&&A^{(1)}_1= i\frac{\cos\theta}{1+\cos\theta}  \\
&& A^{(1)}_2 = -4\cos\theta \frac{\sin^4(\theta/2)}{\sin^3\theta}
\ee
In the limit of small $a \ll 1$, and thus $\theta \ll 1$, one can see that $|A^{(1)}_1|\sim 1/2$, while $|A^{(1)}_2| \sim a/4$  which is much smaller than $|A^{(1)}_1|$. Similarly, the magnitude of higher order harmonics $|A^{(1)}_n|$ with $n>1$ is of the order of $a^{n-1}$. 

For $\sin[2\varphi(t)]$, its Fourier series are quite different. The coefficients can be obtained similarly: 
\be
&&A^{(2)}_1= -\sin\frac{\theta}{2} \frac{\cos\theta}{\cos^3(\theta/2)} \\
&&A^{(2)}_2 = \frac{i}{2} \cos\theta (2\cos\theta-1) \sec^4(\theta/2)
\ee
For small $\theta$, $|A^{(2)}_1| \sim \theta/2$, and $|A^{(2)}_2|\sim 1/2$. It means that $\sin[2\varphi(t)]$ has the largest Fourier component at the second harmonic frequency.  

For the case with $H_D=0$ but $H_K\neq0$, one encounters a differential equation as follows:
\be
\varphi_t= 1+a \sin(2\varphi)
\ee
which can be solved as before by making a variable changes: $\varphi \rar \varphi/2$, and $t\rar t/2$. The solution is given by:
\be
\tan\varphi = \frac{\sin(t')}{\cos(t' + \theta)}
\ee
with $\theta = \arcsin a$ and $t'=\sqrt{1-a^2} t$.  Further,
\be
&&\sin\varphi= \frac{\sin t'}{\sqrt{1-\sin(2t'+\theta) \sin\theta}} \\
&&\sin(2\varphi) =\frac{\sin(2t'+\theta) - \sin\theta}{1-\sin(2t'+\theta) \sin\theta} 
\ee
 In this case, the largest Fourier coefficient of $\sin\varphi$ is at the basic harmonic frequency, and its amplitude is approximately $1/2$ if $\theta$ is small. The magnitude of higher order harmonics is of orders of $o(\theta^n)$.  For $\sin(2\varphi)$, the Fourier coefficients can be obtained analytically, and the largest coefficient is at the second harmonic frequency with amplitude given by $\frac{\cos\theta}{1+\cos\theta} $ which reduces to $1/2$ in the limit of $\theta \rar 0$.
 
 The above analysis shows that in the presence of either $H_K$ or $H_D$, the largest Fourier component of $\sin\varphi$ is always at the basic harmonic frequency while the largest Fourier component  of $\sin(2\varphi)$ is at the second harmonic frequency. Therefore, according the equation of motion of velocity $v(t)$, the ratio of LFM to HFM in the frequency spectrum is a good estimation of the ratio of coefficients of $\sin\varphi$ term to $\sin(2\varphi)$ term, as long as $\alpha H_K \ll H_a$ and $\alpha H_D \ll H_a$. This condition can be easily satisfied by increasing external magnetic field by $H_a$.
 
We can also use perturbation theory to study the case with both $H_K$ and $H_d$ nonzero.  Now we encounter the following equation:
\be
\varphi_t = 1+ a_1 \sin\varphi +a_2 \sin(2\varphi). \label{eqphi3}
\ee
with $a_1\ll 1$ and $a_2\ll 1$. Integrating this equation, we have:
\be
t&&= \int d\varphi \frac{1}{1+a_1\sin\varphi+a_2 \sin(2\varphi)} \nn  \\
&&\sim \int d\varphi (1-a_1\sin\varphi -a_2 \sin(2\varphi))  \nn \\
&&= \varphi + a_1 \cos\varphi + a_2\cos(2\varphi)
\ee
To zeroth order perturbation of $a_1$ and $a_2$, one has simply $\varphi=t$. To first order perturbation, one has: 
\be
\varphi=t-a_1 \cos t-a_2\cos(2t)
\ee
Inserting back to the equation of velocity, one has, up to first order of $a_1$ and $a_2$
\be 
v(t)\sim \gamma' H_a(1+ a_1\sin t + a_2 \sin(2t)).
\ee
Its Fourier coefficients at basic and second harmonic frequencies are $\gamma' H_a a_1$ and $\gamma' H_a a_2$, respectively. The ratio of the two is simply $H_D/H_K$.

\subsection{ Collective coordinate model with noise and disorder}
\begin{figure}[h]
	\centering
	\includegraphics[width=0.5\textwidth]{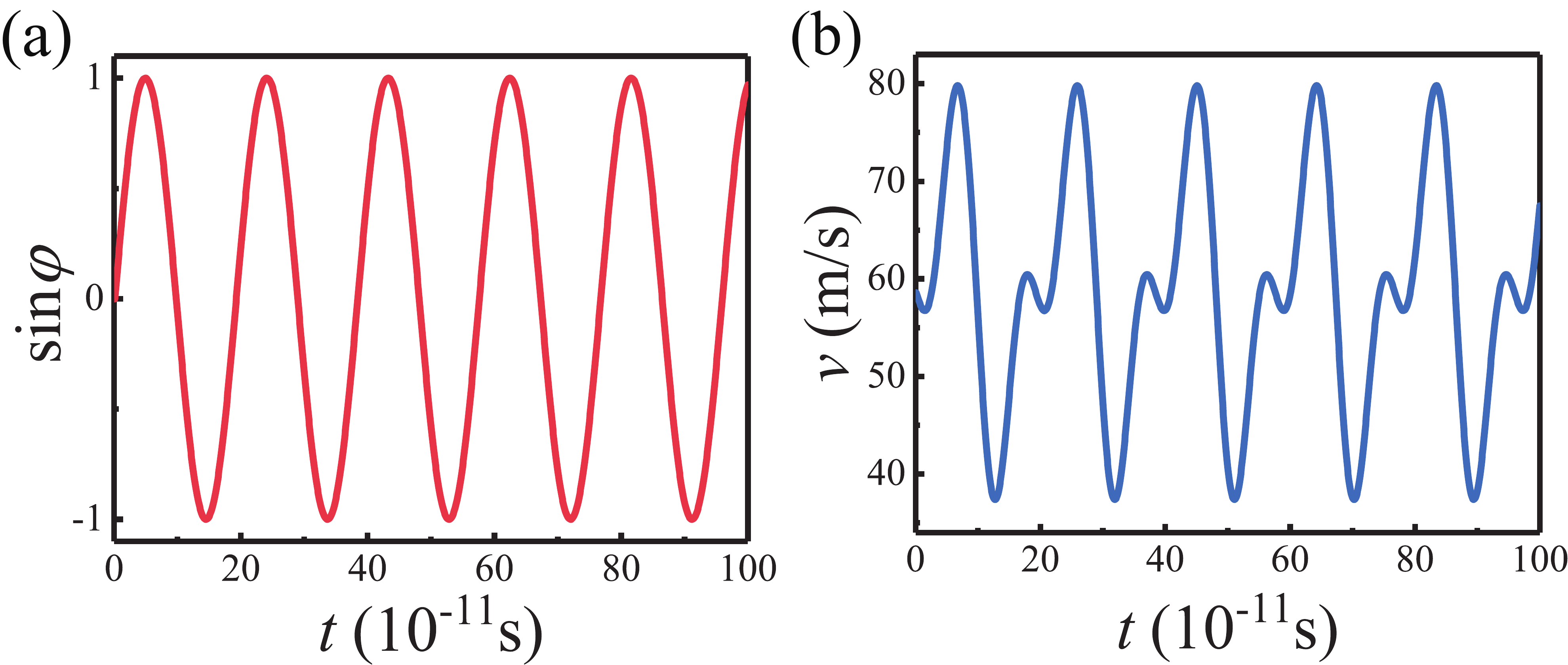}\\
	\caption{Numerical results obtained from CCM. (a) and (b) are $\sin\varphi$ and $v(t)$ varying with time, respectively. Parameters: }
	\label{ccm1}
\end{figure}

In this section, we present more numerical results from the 1D CCM.  In Fig.~\ref{ccm1}, we plot the numerical result of $\sin\varphi$ and $v(t)$ as a function of time, respectively. One can see that $\sin(\varphi(t))$ is almost an sinusoidal, while $v(t)$ shows two different oscillations.

\begin{figure}[h]
	\centering
	\includegraphics[width=0.5\textwidth]{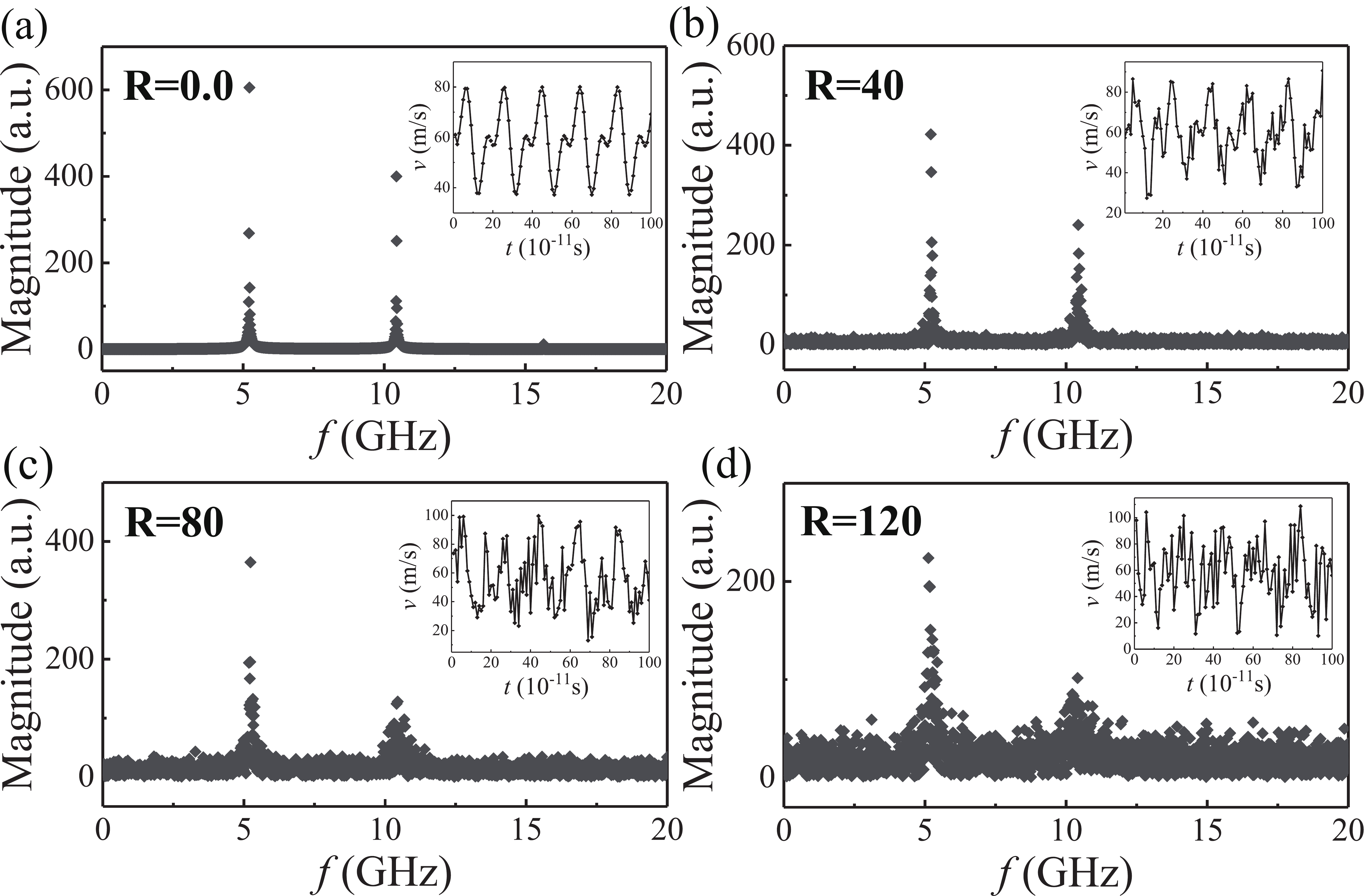}\\
	\caption{The velocity frequency spectrum for different strength of noise in the CCM. The insets are snapshots of time-varying velocity $v(t)$.}
	\label{ccm_noise}
\end{figure}

The effect of external noise can be introduced by the  random field vector $\bf{h}$ on the external field $H_a$ in the CCM. As in the main text,  we adopt a Gaussian noise with strength discribed by $R$. 
 Fig.~\ref{ccm_noise} (a-d) plot the  velocity frequency spectrum for different strength of noise and the insets are snapshots of time-varying velocity $v(t)$. One can see that the velocity is strongly disturbed, and is no longer a well defined sinusoidal. It is therefore hard to tell the information of DW dynamics simply from the profile of velocity. Nevertheless, after a sufficiently long time average, one can still obtain two peaks identified as high-frequency modes (HFM) and low-frequency modes (LFM) in the velocity frequency spectrum, similar to the clean system. 

\begin{figure}[h]
	\centering
	\includegraphics[width=0.5\textwidth]{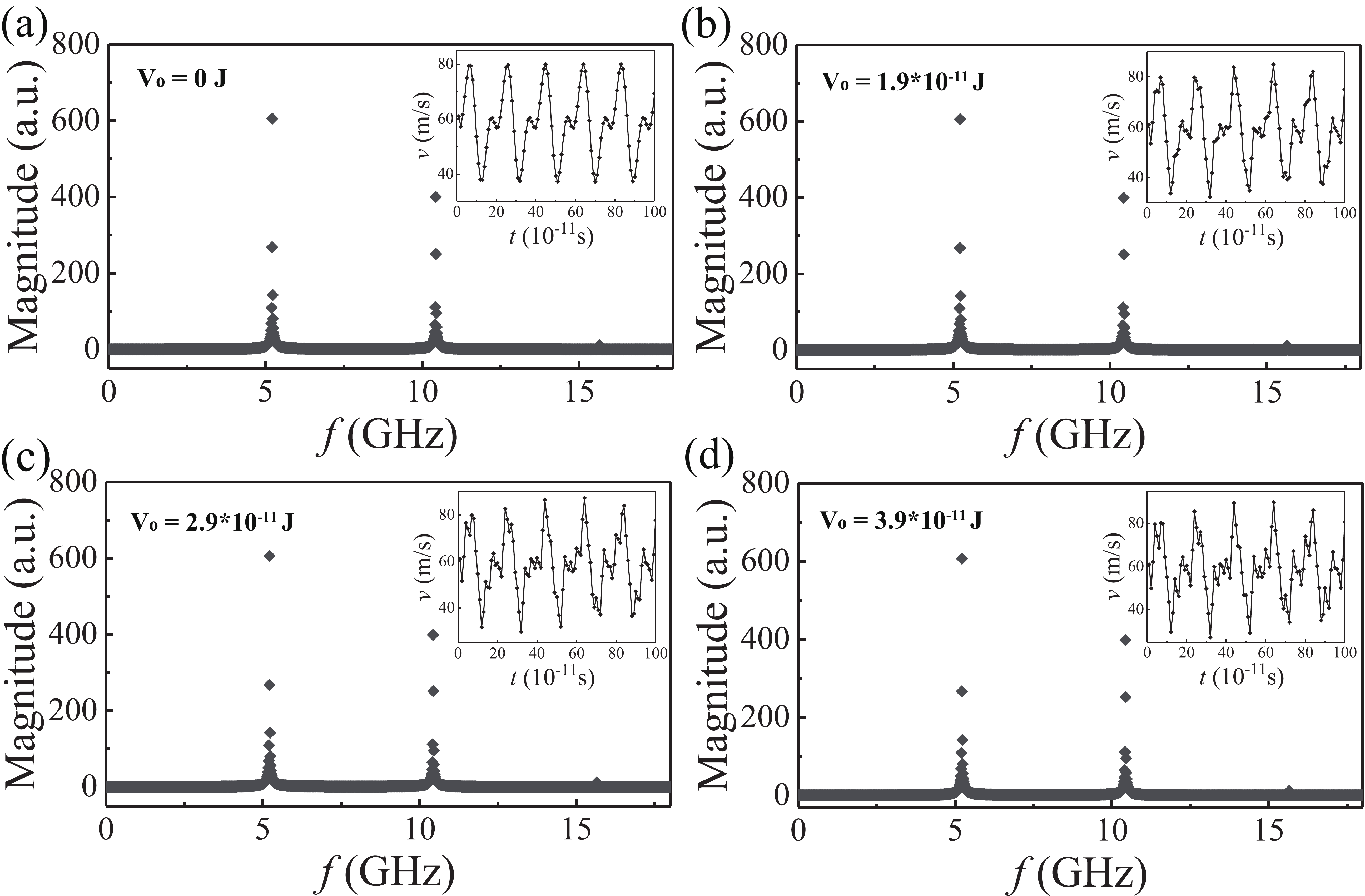}\\
	\caption{The velocity frequency spectrum for different strength of disorder in the CCM. The insets are snapshots of time-varying velocity $v(t)$.}
	\label{ccm_disorder}
\end{figure}

The effect of disorder can be included by introducing a pinning field $H_{\rm pin}(x)$ which depends on the position $x$. The pinning field can be derived from an effective spatially dependent pinning potential $V_{\rm pin}(x)$: 
\be
H_{\rm pin}(x)= -\frac{1}{2\mu_0 M_s L_y L_z} \partial V_{\rm pin}/\partial x
\ee
The pinning potential can be chosen to be periodic with strength described by $V_0$: 
\be
V_{\rm pin} (x) = V_0 \sin\Big(\frac{\pi x}{p}\Big),
\ee
where $p$ is the spatial period. 
Fig.~\ref{ccm_disorder} (a-d) plot the  velocity frequency spectrum for different strength of disorder and the insets are snapshots of time-varying velocity $v(t)$. 


\subsection{Domain wall dynamics driven by electric current}
In ferromagnetic nanowires, the DW can also be driven by spin-polarized electrical currents due to the spin transfer torque (STT) exerted on the local magnetization. Including the STT, the LLG equation can be written as \cite{Boulle2013}: 
\be
\dfrac{\partial \bm{ {m}}}{\partial t}&&=-\gamma \bm{ {m}} \times {\bm H}_{\rm eff}+\alpha \bm{ {m}} \times\dfrac{\partial \bm{ {m}}}{\partial t}  \nn \\
&& + b_J (\hat{\bm j}\cdot\nabla){\bm m} -c_J {\bm m}\times (\hat{\bm j}\cdot \nabla){\bm m}.
\ee
The first and second term have been introduced in the main text. The third term is the adiabatic STT $b_J=P \mu_B j/eM_s$ with $e$ being the electron charge, P the spin polarization, $\mu_B$ the Bohr magneton, $j$ the magnitude of current and $M_s$ the saturation magnetization. Here, $\hat{\bm j}$ is the unit vector of the local current density. The fourth term is the non-adiabatic STT with $c_J$ being the magnitude of nonadiabticity. Usually one introduces a dimensionless parameter $\xi = c_J/b_J$ to represent the nonadiabaticity. 

\begin{figure}[h]
	\centering
	\includegraphics[width=0.5\textwidth]{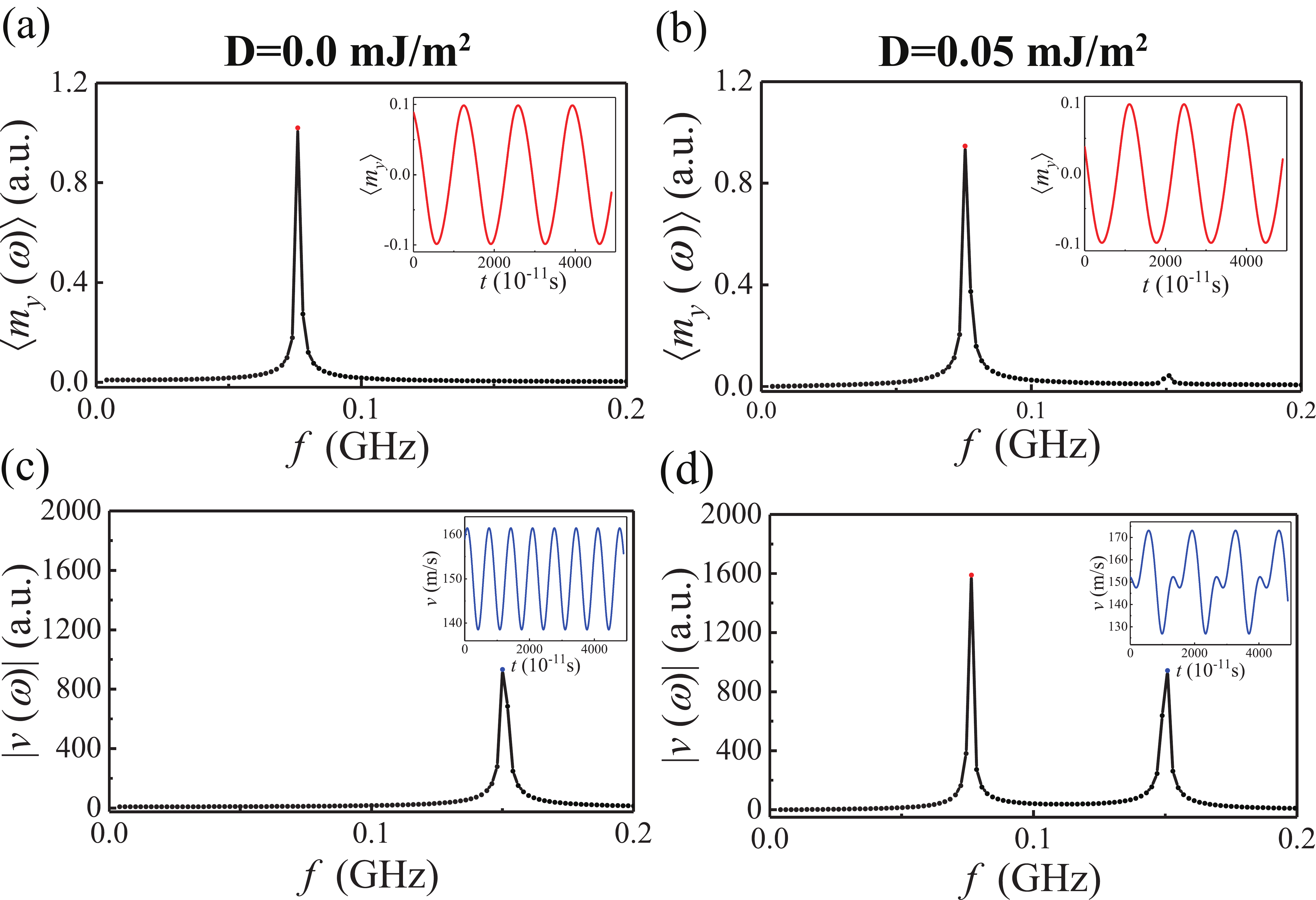}\\
	\caption{(a) and (b) Frequency spectra of  $\langle m_{y}(t) \rangle$ without and with DMI, respectively.  (c) and (d) Frequency spectra of  $v(t)$ without and with DMI, respectively.  The red and blue points correspond to the LFM and HFM peaks respectively.	Inset: $\langle m_{y} \rangle (t)$ (red dotted line) and $v(t)$ (blue dotted line) signals. Electric current $j=3\times 10^{12}$A/m$^2$, spin polarization $P=0.8$ and nonadiabaticity $\xi=0.04$.}
	\label{figcurrent1}
\end{figure}

In Fig.~\ref{figcurrent1}, we plot the frequency spectrum of  $\langle m_{y}(t) \rangle$ and $v(t)$ for the case with and without DMI.  In the precessional regime with zero DMI strength $D=0$, one observes that, both the $\langle m_{x,y}(t) \rangle$ and $v(t)$  oscillate periodically with time, as shown in the inset of Fig.~\ref{figcurrent1}. However, they have different oscillating frequency. 
 In Fig.~\ref{figcurrent1}(a) and (c), one plots the magnitude of Fourier transform $|\la m_y(\omega)\ra$ and  $|v(\omega)|$ for the case of $D=0$. It is shown that, there is only one single peak in the frequency spectrum of $m_y(t)$ and $v(t)$. However, the peak frequency of velocity is doubled compared with  that of $m_y$. 
 
 In the presence of  DMI, the time-varying signal $\langle m_{y}(t) \rangle$ is almost the same as  $D$ = 0, and the oscillation frequency remains unchanged (inset of Fig.~\ref{figcurrent1}(b)).  However, the behavior of velocity signal $v(t)$ is radically changed, as shown in the inset of Fig.~\ref{figcurrent1}(d).   
 Figures ~\ref{figcurrent1}(b) and (d) plot  the  frequency spectra of $\langle m_{y} \rangle$ and $v(t)$ with nonzero DMI . There  emerges two spectral peaks with  different magnitudes in $|v(\omega)|$.

\begin{figure}[h]
	\centering
	\includegraphics[width=0.5\textwidth]{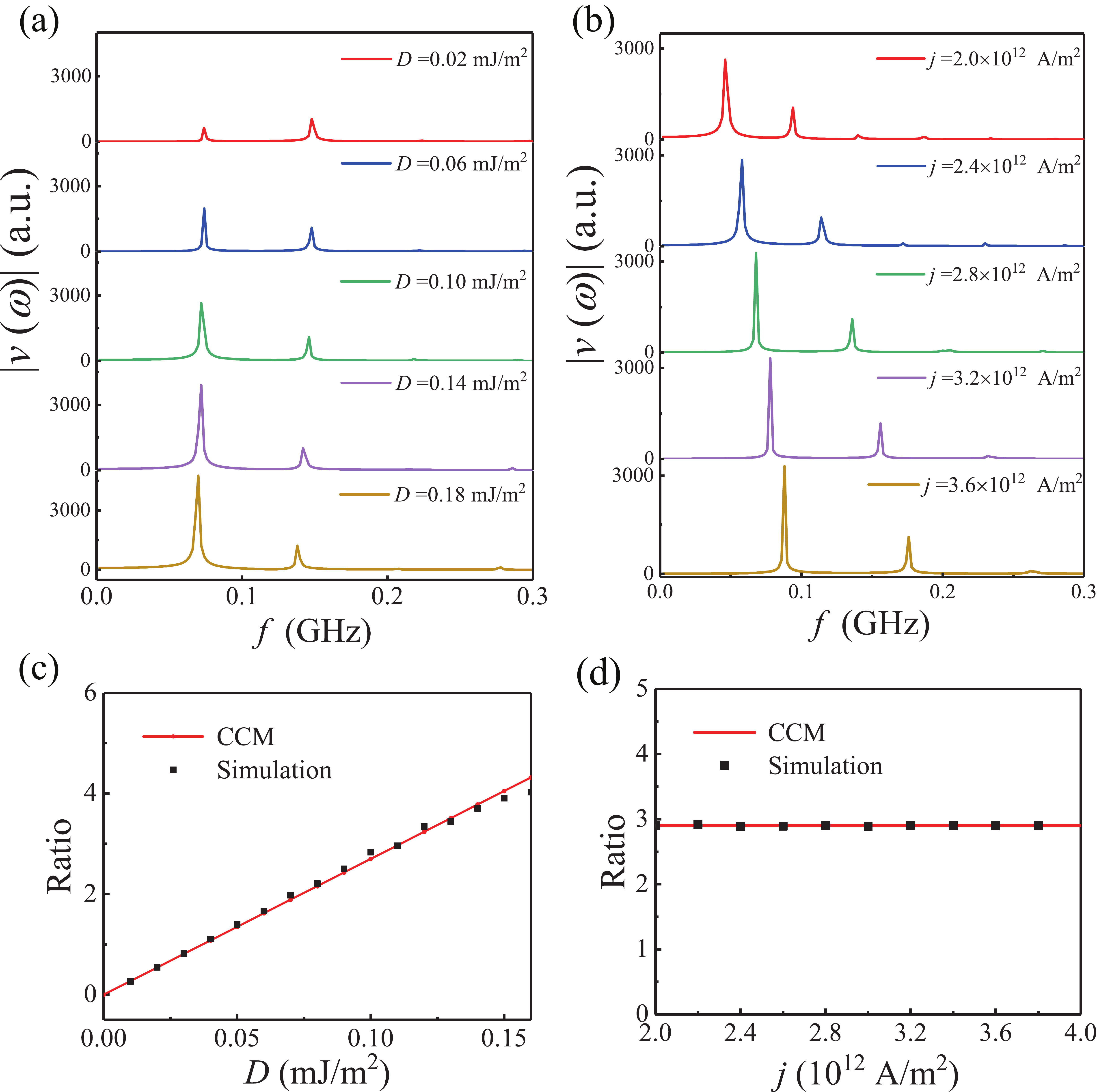}\\
	\caption{Frequency spectrum of DW velocity $v(t)$ for different values of (a)  $D$ and (b)  current $j$. Dependence of the ratio of  LFM to HFm  on  $D$ (c) and $J$ (d). 
		Squares are results from micromagnetic simulation and the solid line  from the  CCM. Parameters: $P=0.8$, $\xi=0.04$, $j=3\times 10^{12}$A/m$^2$ (a) and $D=0.1$mJ/m$^2$ (b).}
	\label{figcurrent2}
\end{figure}

We further investigate the influence of DMI strength $D$ on the two emerging oscillating peaks in the velocity frequency spectrum.  We first  fix  the magnitude of electric current $J$ and gradually increase the DMI strength $D$. As shown in Fig.~\ref{figcurrent2}(a), there are always two peaks in the velocity frequency spectrum with the frequencies keeping unchanged. However, with increasing DMI strength, the magnitude of the LFM peak increases, while the magnitude of HFM peak remains  unchanged. If we extract the LFM/HFM ratio from the frequency spectral data and plot them as a function of DMI intensity  (Fig.~\ref{figcurrent2}(c)), one can see that the ratio linearly increases with the DMI strength $D$. 
We also  study the velocity frequency spectrum  under different electric current $J$ when the DMI strength is fixed, as shown in Fig.~\ref{figcurrent2}(b) and (d). Even though the oscillating frequency increases with increasing $J$, the magnitude ratio of LFM/HMF remains unchanged as long as the DMI strength $D$ is fixed.  

In the current-driven case, one can also introduce the collective-coordinate model as follows: 
\be
	&&\dfrac{d\varphi}{dt}+\alpha\dfrac{1}{\Delta }\dfrac{dq}{dt}=\frac{c_J}{\Delta}, \label{eqphi} \\
		&&\alpha \dfrac{d\varphi}{dt} - \dfrac{1}{\Delta }\dfrac{dq}{dt} =\frac{b_J}{\Delta}+\gamma H_{K} \sin 2 \varphi - \gamma H_{D} \sin \varphi, \label{eqq}
		\ee
Compared with the magnetic field driven case, the difference is the substitution of $\gamma H_a$ with $c_J/\Delta$ and an additional term of $b_J/\Delta$. It is shown in Fig.~\ref{figcurrent2} that the results of CCM agree well with that of micromagnetic simulation.

\end{document}